\documentclass{fundam}

\usepackage[T2A]{fontenc}
\usepackage[utf8]{inputenc}
\usepackage[english]{babel}

\usepackage{latexsym}
\usepackage{amssymb}
\usepackage{listings, chngcntr} 
\usepackage{color}
\usepackage{xcolor}
\usepackage{soul}  
\usepackage{graphicx}
\usepackage{listings, chngcntr} 

\publyear{22}
\papernumber{2102}
\volume{185}
\issue{1}

\usepackage{url} % takes care of hyperlinks, preferred over hyperref
\usepackage[ruled,lined]{algorithm2e}% provides Algorithm environment
\usepackage{tikz}
\usetikzlibrary{automata, positioning, arrows}

\definecolor{mygreen}{rgb}{0.2,0.6,0.2}

\lstdefinestyle{cplusplus}
{ 
	backgroundcolor=\color{white},   
	basicstyle=\linespread{1}\fontsize{10pt}{10pt}\selectfont\ttfamily,            
	breakatwhitespace=false,         
	breaklines=true,                 
	captionpos=b,                    
	commentstyle=\color{mygreen},    
	deletekeywords={...},            
	escapeinside={\%*}{*)},          
	extendedchars=true,              
	keepspaces=true,                 
	keywordstyle=\color{blue},       
	language=C,                     
	morekeywords={__kernel, __local, __global, barrier}, 
	rulecolor=\color{black},         
	showspaces=false,                
	showstringspaces=false,          
	showtabs=false,                  
	stepnumber=1,                    
	stringstyle=\color{mymauve},     
	tabsize=1,
}

\lstdefinestyle{pml}
{ 
	backgroundcolor=\color{white},   
	basicstyle=\linespread{1}\fontsize{10pt}{10pt}\selectfont\ttfamily,            
	breakatwhitespace=false,         
	breaklines=true,                 
	captionpos=b,                    
	commentstyle=\color{mygreen},    
	deletekeywords={...},            
	escapeinside={\%*}{*)},          
	extendedchars=true,              
	keepspaces=true,                 
	keywordstyle=\color{blue},       
	language=C,                     
	morekeywords={chan, byte, od, fi, atomic, proctype, select, run, active, inline}, 
	rulecolor=\color{black},         
	showspaces=false,                
	showstringspaces=false,          
	showtabs=false,                  
	stepnumber=1,                    
	stringstyle=\color{mymauve},     
	tabsize=1,
}

\begin{document}

\title{Auto-Tuning High-Performance Programs\\ Using Model Checking in Promela}

\address{garanina{@}iis.nsk.su}

\author{Natalia Garanina\thanks{Natalia Garanina was supported by the scholarship of DAAD (German Academic Exchange Service).}\\
On visit at the University of Muenster, Germany 
\and Sergey Staroletov
\and Sergei Gorlatch\thanks{Sergei Gorlatch is supported
by the DFG project Nr. 470527619 (PPP-DL) at the University of Munster.
}\\
University of Muenster, Muenster, Germany } 

\maketitle

\runninghead{N. Garanina, S. Staroletov, S. Gorlatch}{Auto-Tuning Using Model Checking}

\begin{abstract}
The paper combines research approaches that traditionally have been disjoint: 1) model checking as used in formal verification of programs, and 2) auto-tuning as often used in high-performance computing. 
Auto-tuning frameworks optimize parallel programs by finding the optimal values of the performance-critical parameters --- so-called tuning parameters --- for a particular high-performance architecture and input data size.  As there are many parameters influencing program's performance, finding the optimal parameter configuration is a hardly manageable task even for experts.
Auto-tuning automates this process, but it is often time-consuming. We apply model checking for accelerating auto-tuning by using a counterexample constructed during the verification of the optimality property of the program. We describe in detail an implementation of our approach for programs written in OpenCL -- the standard for programming modern high-performance architectures -- using the model representation language Promela and the popular SPIN verification tool, and we report experimental results for an application use case. 
\end{abstract}

\begin{keywords}
model checking, temporal logics, counterexamples, high-performance computing, auto-tuning, SPIN, Promela
\end{keywords}

\section{Motivation and Related Work} \label{intro}

We aim at bringing together two research areas that have traditionally been disjoint:\linebreak 1) model checking used in formal verification of programs, and 2) auto-tuning used in the optimization of programs for modern high-performance architectures that comprise multi-core universal processors (CPU) and many-core Graphics Processing Units (GPU).

\emph{Auto-tuning} \cite{hoos2011automated} is an important case of the general concept of automated algorithm configuration and parameter tuning.
It has become especially popular recently in developing parallel programs: it finds the values of the performance-critical program parameters that ensure optimal performance of the program code on a particular target architecture and problem data size. Typical examples of such tuning parameters for parallel programs in the OpenCL language are: number of threads, tile size, workgroup size, shared-memory block size, etc.
The potentially large number of tuning parameters and their complicated interplay in influencing program performance makes it very difficult even for an expert to find a (near to) optimal combination of parameter values that ensures good performance on a particular machine and for a particular size of input data.

For the implementation of auto-tuning, an auto-tuning framework (\textit{auto-tuner}) is used: the framework automatically generates a search space of parameter configurations and then examines this space by running the program for chosen configurations on a real hardware of the target high-performance system. The currently popular auto-tuners include OpenTuner \cite{OpenTuner}, PATUS \cite{PATUS}, ATLAS \cite{ATLAS}, FFTW \cite{FFTW}, CLTune \cite{CLTune}, ATF \cite{ATF}, to name a few. They use a variety of techniques to find the optimal solution, including simulated annealing, random search, genetic algorithms, machine learning, and dynamic programming. Only few of auto-tuners nowadays offer exhaustive search, which guarantees to find the optimal result but implies extremely high time costs for checking all possible configurations: the search for a solution of high quality may take several hours or even days on a high-performance system.

In this paper, we attempt to improve auto-tuning by using modern model checking tools. The goal of our approach is to automatically find good tuning parameter values faster than existing auto-tuners, and without using real hardware. Model checking is widely used for formal program verification: it checks the satisfiability of a particular program property for all scenarios of a program model behavior. Both the model and the property are specified using formal languages. For example, a program model can be represented as a Kripke structure, and a property as an LTL temporal logic formula. If the property is not satisfied then the model checking method can find a \textit{counterexample}, i.e., conditions and sequence of actions of the model that imply that the property is not satisfied. 

Our approach to auto-tuning based on model checking relies on exhaustive or swarm search by using a counterexample constructed during the verification of the optimality property for a program which is to be auto-tuned. We have been inspired by previous research on using counterexamples for solving optimization problems and, in particular, compiling optimal schedules for reactive systems \cite{BrinksmaMader02,Malik18,RuysBrinksma98,Ruys03,Wijs09}. A special edition of Dagstuhl seminar \cite{Dagshtul} was devoted to the interrelation of automatic planning/optimization and model checking methods. The recent progress in model checking has achieved that modern tools examine each element of the search space using various shortcuts, such as symbolic encoding, partial order reduction, and abstraction \cite{HBM}. Due to these shortcuts, exhaustive exploration of larger search spaces can be done in a shorter time. 

An important expected benefit of using model checking for auto-tuning is that the target high-performance architecture does not have to be available when searching for the optimal parameter values of a parallel program designed for this architecture. We are not aware of previous approaches to use model checking for auto-tuning programs designed for high-performance systems with modern multi- and many-core processors. Malik and Pena \cite{lazreg2019multifaceted} use a custom model checker to compute optimal parameter configuration for multi-variable systems in the automotive industry. This task can be considered similar to auto-tuning, but it does not use the counterexample method that we apply in the present paper. 

In this paper, the target area for auto-tuning are parallel programs written in OpenCL~\cite{OpenCL} which is an open standard approach to programming various modern architectures: multi-core CPU, many-core GPU, FPGA, and others. In order to ensure generality and flexibility of our approach, we assume the abstract platform model of OpenCL as our target architecture, rather than a particular kind of architecture. As our model checking tool, we use the SPIN verifier \cite{SPIN} with its modeling language Promela \cite{promelagr} whose semantics combines the CSP (Communicating Sequential Processes) \cite{Hoar85} parallelism and the actor model \cite{GaspariZavattaro99}. 

We initially proposed our idea of using model checking for auto-tuning in the short, Russian-language paper \cite{garanina2021autotuning} where we model the Nvidia's Fermi architecture as an abstract multiprocessor with a warp scheduling for processing elements. The conference paper \cite{garanina2022model} refines our initial ideas and extends them by developing a first proof-of-concept implementation of auto-tuning for OpenCL programs using the SPIN model checker and conducting first experiments with it. This paper is a substantially rewritten and extended version of \cite{garanina2022model}, with the following major additions: 
1) we describe particular details of our approach, in particular we formally define the Promela processes for devices, units and the service clock, and 
2) we significantly extend the scope of our experiments in that we describe a particular application of our approach to auto-tuning an OpenCL program for a representative example of computing problems based on reduction, namely, for  computing the minimum in integer array. 

The structure of this paper and its main contributions are as follows. Section~\ref{optmch} presents the idea of our approach as an adaptation of model checking and the technique of counterexamples to auto-tuning high-performance programs. Section~\ref{prll} describes the standard platform model of OpenCL and how our abstract model of the OpenCL platform can be represented in the Promela language. Section~\ref{spin} presents the use of the SPIN model checking tool for auto-tuning, i.e., finding the optimal parameter configuration of an OpenCL program. Section~\ref{limited} discusses how we can adapt our general auto-tuning approach for the case of limited computation resources. Section~\ref{exprm} reports our extensive experiments with particular OpenCL programs and their results. In Section~\ref{apps}, we apply our model checking based approach to auto-tuning for a program that computes the minimal value in an integer array in parallel, and we compare results of our approach with the real execution of the program.
Section~\ref{conc} concludes our presentation, summarizes our findings, and provides an outline for future research.

\section{Model Checking for Auto-Tuning: The Idea of Our Approach} \label{optmch}

Our main idea in adapting model checking to auto-tuning high-performance programs is based on using the general counter\-example-guided approach as described, e.g., in \cite{BrinksmaMader02}. We target multi-threaded programs that transform some input data to output data in a highly parallel manner. The task of auto-tuning such parallel programs is to find the values of performance-critical program parameters (we call them \textit{tuning parameters}) that enable obtaining the result of the program in the minimum possible time. 

We start outlining our approach by formulating a general optimality property as \textit{the over-time property $\Phi_o$} for the impossibility of program termination within a particular time interval as follows: ``A parallel program \textit{cannot} terminate within $T$ units of time'', i.e., the program requires a run time that is longer than $T$. To formalize this property, we can use a temporal logic, such that the over-time property can be formally verified in the model that represents the execution of the parallel program on a particular target architecture, using some model checking tool. If this verification shows that the property is not satisfied, this means that the computation can indeed terminate within the model time $T$. In this case, the model checker constructs a counterexample that describes the conditions under which the program terminates within a model time not greater than $T$, and also the corresponding configuration of tuning parameters that are responsible for the achieved program's performance. Next, we can gradually approximate the given termination model time $T$ and check the over-time property again and again, until we reach a \textit{minimal} model time $T_{min}$ of the program: if we decrease this time by only one unit of time, the model checker will prove that the $\Phi_o$ property is satisfied, i.e., the program really cannot terminate within $T_{min}-1$ units of time. This way, we find both the minimal model run time of the parallel program and the corresponding values of tuning parameters to achieve this run time; in other words, we have auto-tuned our program.

\begin{figure}[ht]
	\begin{center}
	\includegraphics[ height=180pt]{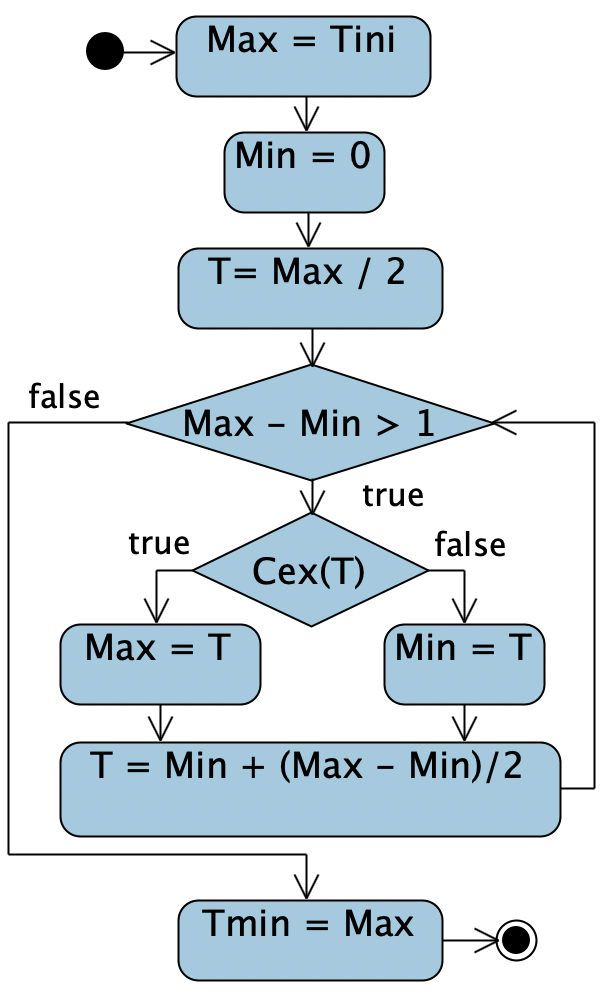}
	\end{center}
	\vspace{-4mm}
	\caption{The bisection method for finding a minimal termination time}\label{Bisect}
\end{figure}

For implementing this idea of auto-tuning using model checking, we should choose a suitable initial time value $T_{ini}$ and determine a procedure for moving from $T_{ini}$ towards the minimal model run time $T_{min}$. To find a minimal termination model time, we may use, for example, the bisection method shown in Fig.~\ref{Bisect}: predicate $C_{ex}(T)$ is true iff a model checker generates a counterexample for the model and $\Phi_o$ property with time $T$. The initial value of the termination model time $T_{ini}$ can be specified by simulating the program model: usually model checking tools allow for simulating model execution scenarios, such that it is possible to reveal the value of parallel program execution time in any of the scenarios. 

Summarizing, our proposed method of using model checking with counterexamples for auto-tuning a parallel program proceeds in four steps, as follows:
\begin{enumerate}
\item Represent the parallel program with its tuning parameters and target architecture in the language (in our case Promela) of a model checking tool (in our case - SPIN).
\item Formulate the over-time property $\Phi_o$ for the inability to terminate the program execution within time $T$ in the language of a model checking tool.
\item Search for the minimal program termination time $T_{min}$ starting from an initial time $T_{ini}$ provided by a simulation of the model constructed in Step 1.
\item Extract information about the optimal configuration of tuning parameters from the counterexample found for the minimal time $T_{min}$.
\end{enumerate}

\section{Our Modeling Approach: From OpenCL to Promela} \label{prll}

In this section, we describe how our auto-tuning method outlined in the previous section can be implemented for our particular use case -- auto-tuning of OpenCL programs using the formal modeling language Promela and the verification tool SPIN for this language. We target OpenCL, because it covers programming for practically all currently used and emerging parallel architectures, including: universal multi-core processors (CPU), many-core Graphics Processing Units (GPU), FPGA, etc. 

\subsection{An Abstract OpenCL Platform Model} \label{aproc}

\begin{figure}[ht]
	\begin{center}
	\includegraphics[width=0.6\linewidth]{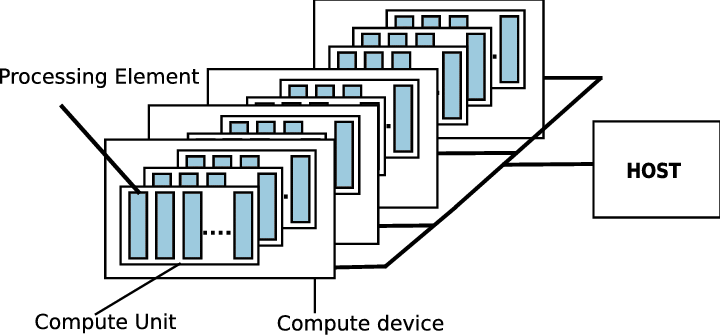}
	\end{center}
	\caption{Entities in OpenCL platform \cite{OpenCL}}\label{HostDevice}
\end{figure}

Figure~\ref{HostDevice} shows the components of our architecture \texttt{Abstract Platform} based on the so-called platform model as defined in the OpenCL standard~\cite{OpenCL}. 

The abstract model in Fig.~\ref{HostDevice} corresponds to the currently most widespread use case of OpenCL: \textit{compute devices} represent (possibly multiple) GPUs that are connected to the host CPU. The host is responsible for loading data to and from the \textit{global memory} of devices. Every device includes \textit{compute units} to process the data, while each unit contains \textit{processing elements}. For example, the newest Ampere A100 architecture of Nvidia Corp.\ comprises in one device (GPU) altogether 108 compute units (called streaming multiprocessors in the Nvidia terminology) with a total number of 6192 processing elements (so-called CUDA cores). 

For simplicity of modeling, we assume in our model that every abstract compute unit includes $2^n$ abstract processing elements that are universal calculators.
A compute unit has fast \textit{local memory}, to which all its processing elements have access. Data from global memory of a device can be loaded there. The difference of the access speed to local and global memory is usually between one and two orders of magnitude, depending on a particular processor. 

The Promela language allows us to model all components of our abstract platform model -- devices, units, processing elements, and the host process that distributes computations among devices -- using different Promela processes. These processes are hierarchically linked: the host process starts devices while devices activate their units that regulate the work of their processing elements. The processes synchronize their work by using handshake message channels for launch and finish commands, as well as for reporting termination. We model the difference in the speed of access to local and global memory by using constant values that specify the memory access time. We simulate the global time in the system by a process counting the number of active processing elements to determine the moment when to increase the time. We do not take into account the communication time between platform components, which in the practice of GPU programming is significantly lower than other time delays. We describe the details of Promela implementation of the abstract platform in Section~\ref{spin}.

\subsection{An Abstract OpenCL Program}

Every OpenCL program consists of two logical parts: a \textit{host program} and a \textit{kernel}: the host program is executed on a CPU, and the kernel is executed in parallel by all processing elements of the device (e.g., a GPU) connected to the host. 

Listing~\ref{host_src} shows the structure of the host program: the host compiles the kernel program at run time and organizes its work with data, in particular reserving the global device memory, copying input data into it, and uploading the result data from this memory into the host memory for subsequent processing. The device (usually GPU) performs parallel data processing by executing the kernel code simultaneously on multiple processing elements.

\bigskip

\counterwithout{lstlisting}{section}
\begin{lstlisting}[label={host_src},caption={Host program in OpenCL: abstract structure}, style=cplusplus, numbers=left]
int main (void) {
  Initialize OpenCL
  Compile kernel source code
  Reserve global memory on device, copy input data to device
  Execute kernel (instances)
  Copy output data from the device
  Analyze and process output data
}
\end{lstlisting}

Usually, an OpenCL program processes input arrays and outputs the results of that processing. Therefore, kernels evaluate expressions in the parallel manner, based on the available data, depending on the array index, and assign the results of the calculation to separate elements of the output array. The involved arrays are often multidimensional, with corresponding multidimensional indices.

Listing~\ref{kernel} shows a typical structure of an OpenCL kernel; each instance of the OpenCL kernel is executed by a so-called \textit{work item} that has an identifier corresponding to the array index. Thus, work items perform the same calculations for different array elements as it is usual in the concept of data (or SIMD) parallelism \cite{subhlok1993exploiting}. If the computations of an iteration of the loop as in Listing~\ref{kernel} depend on the computations of another iteration, i.e., on the array index, then OpenCL-provided \textit{barriers} for synchronizing computations are used. When executed on the underlying architecture, several work items are united in a \textit{workgroup}, either explicitly by the programmer or automatically by a specific OpenCL implementation, such that all work items of a workgroup are executed on the same compute unit of the device.

\begin{lstlisting}[label={kernel},caption={Typical scheme of an OpenCL kernel program}, style=cplusplus, numbers=left]
__kernel void abstract_kernel (__global float* N_g, __global float* R_g, int size) {
  __local float N_l[TS];
  int idx_g = get_global_id(0); 
  int idx_l = get_local_id(0); 
  float result = 0;
  for (int i = 0; i < size / TS; ++i) {
    // access to global memory
    N_l[idx_l] = f(i, idx_l, size, N_g); 
    // waiting for local co-workers
    barrier (CLK_LOCAL_MEM_FENCE); 
    if (b(idx_l)) // access to local memory
       for (int k = 0; k < TS; ++k) 
          result = g1(k, idx_l, N_l, result);
    else for (int k = 0; k < TS; ++k) 
          result = g2(k, idx_l, N_l, result);
    // waiting for local co-workers
    barrier (CLK_LOCAL_MEM_FENCE);
  } 
  // copy the result of this item to global memory
  R_g[h(idx_g, size)] = result;
}
\end{lstlisting}

In the OpenCL language (which is syntactically based on C), there are four memory types: \textit{global, constant, local, and private}. All work items and the host have access to global memory. Constant memory is an immutable part of global memory. Elements of one workgroup have access to local memory. Work items can also use their private memory. Since the access to local memory on the device is significantly faster than access to global memory, the kernel often specifies which chunks of data (so-called \textit{tiles}) and how are loaded into local memory. 

\begin{figure}[ht]
	\begin{center}
	\includegraphics[width=0.6\linewidth]{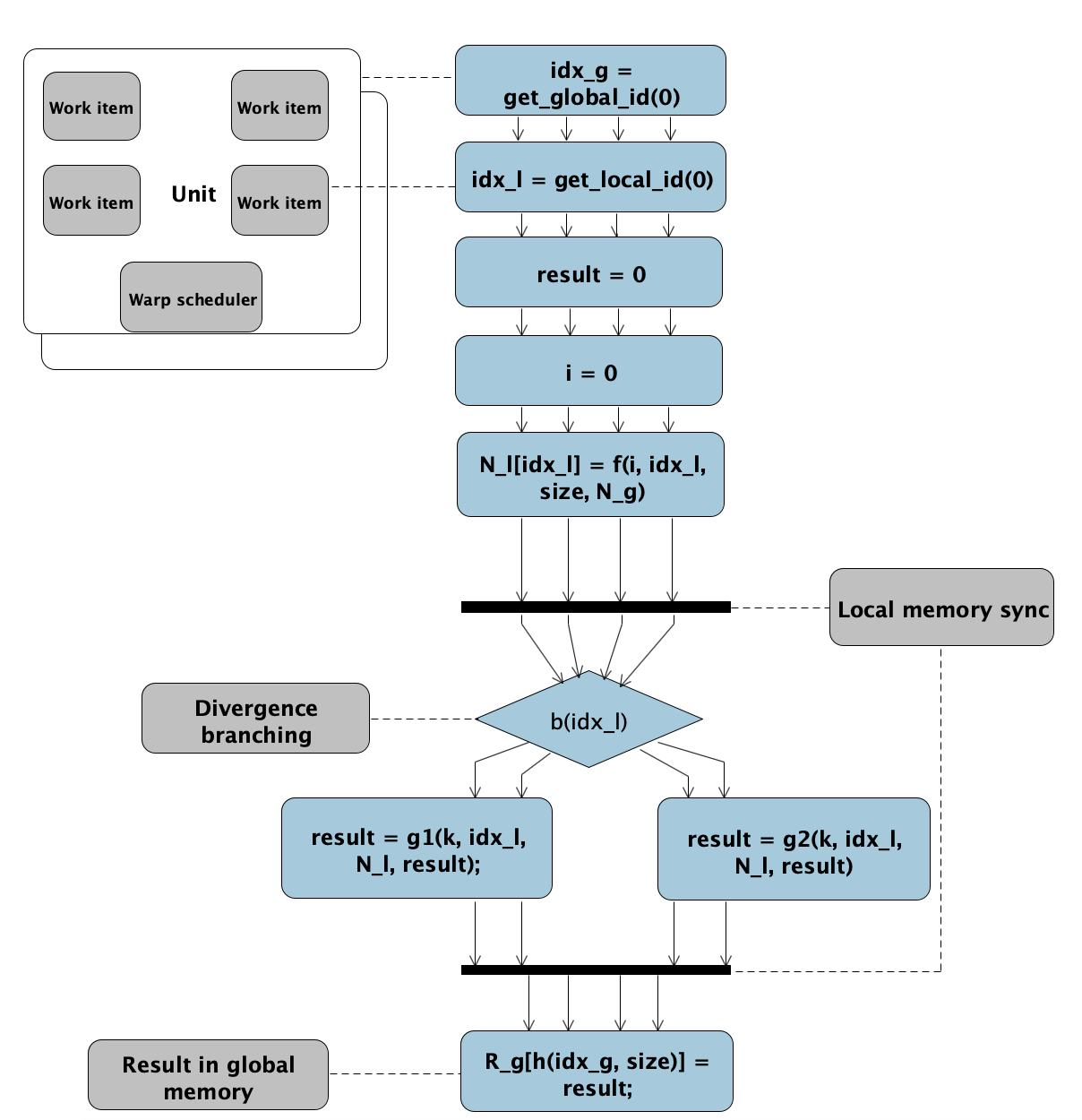}
	\end{center}
	\caption{Operation of a typical OpenCL kernel }\label{AbsScheme}
\end{figure}

In the following, we illustrate how we can model the execution of OpenCL programs using the model checking tools in general and the Promela language in particular. For that, we define the following abstract representation of an OpenCL kernel program which transforms an input array to resulting output array by performing abstract computation with four abstract functions over integer and floating point variables.

Let us describe how the program in Listing~\ref{kernel} performs data-parallel computations on the elements of arrays. For clarity, the scheme of program operation is also shown in Fig.~\ref{AbsScheme}.  For each index of an array, a separate work item calculates the local result of the kernel code execution. In our example program in the listing, the input data is array \texttt{N\_g} of size \texttt{size}. The output (result) here are the elements of the \texttt{R\_g} array. To reduce accesses to global memory, line 2 in the listing declares a local array \texttt{N\_l} of size \texttt{TS} whose elements depend on the input data. Each member of a workgroup that obtains the value of its index \texttt{idx\_l} in line 4, applies function \texttt{f} to calculate its value in parallel using local array \texttt{N\_l[idx\_l]} in line 8; this value can be used by all members of the workgroup, so in line 10 every workgroup member waits until the local data of all members of this workgroup are processed. This waiting is implemented by the synchronization operator \texttt{barrier}. Further in lines 12-14, the iterative calculation of the \texttt{result} depends only on the local data and the index of the group element. In addition, depending on the index, these calculations can be performed in different ways: in line 11, boolean function \texttt{b(idx\_l)} controls the calculation options (function \texttt{g1} or \texttt{g2}). In line 17, the work item waits for the completion of the calculations of all co-workers with the current local data, and at the next iteration of the loop in line 6, local data are updated based on the next portion of global data. When input global data are fully processed, the result of the work item is saved in global memory in array element \texttt{R\_g}, the index of which depends on the size of the input data and on the \texttt{idx\_g} global index received by the work item in line 3.

Listing~\ref{kernel} is a deliberately simplified version of an OpenCL kernel. In general, arrays may be multidimensional; there may be several input arrays, and, in addition, there may be other input variables and constants located in global memory; the output can comprise multiple arrays; there are no changes in global memory that require synchronization of all processes in all workgroups. Of course, the pseudocode in Listing~\ref{kernel} can be easily enriched with these details without going beyond the scope of its functions and operators and, thus, our approach is applicable in the general case.

Our Promela-based approach relies on the fact that Promela, like OpenCL, is close to the C language. Hence, we can translate OpenCL programs into the Promela language taking into account the following restrictions. As the SPIN verification framework assumes an abstraction from the computational aspects of programs and is intended to check the interaction, synchronization and coordination of parallel processes, all data in the Promela model must be of a finite type. Moreover, when modeling the abstract OpenCL kernel in Promela, we abstract from specific calculations. To find the optimal parameter values, we only take into account the time of these calculations, which depends on the number of accesses to the global and local memory and the ratio of them. Therefore, computations in lines 8, 12, 14, and 20 of Listing~\ref{kernel} are replaced in the Promela model of the kernel by the code that implements time ticking required for these computations. For the same reason, we ignore the parameters and local variables of the kernel that specify the content of the data for calculations (lines 1-5). This abstraction can be refined for a particular program and hardware by estimating run time for every function used in the program: we take into account the time of all primitive operations in the function and the ratio of the global memory access time to the local memory access time (these characteristics are usually known from the hardware specifications). However, the amount of these calculations which depends on \texttt{size} must be taken into account too. Hence, the code in Listing~\ref{kernel} is representative enough up to the number of loops and program control structures, because for modeling massive parallel computations, only the size of the input data matters.

An important aspect that we take into account in our Promela-based modeling is the synchronization of workgroup elements relative to changes in local and global memory. To provide local synchronization, our model introduces barrier processes that are responsible for separate workgroups, while for global synchronization we introduce a barrier process for all work items. In our example kernel, there is no global barrier, but its implementation in Promela is quite similar to the implementation of the local barrier. The role of the host program in our Promela modeling is reduced to the process of ``compiling'' the kernel, i.e., distribution of computations on a given device. We present more details of the Promela implementation of the execution of both OpenCL kernel and its host program in the next section, which also describes the application of model checking to optimize the performance of OpenCL programs.

\section{Using SPIN for Auto-Tuning OpenCL Programs} \label{spin}

This section describes in detail how we implement the four steps of our counterexample-based approach (Section~\ref{optmch}) in order to auto-tune OpenCL programs using the model-checking tool SPIN and its language Promela. Our modeling of OpenCL programs execution is based on the standard OpenCL semantics \cite{OpenCL}.

The OpenCL semantics prescribe that the OpenCL compiler allocates a host (CPU) connected to a device (CPU, GPU, or FPGA) for the execution of an OpenCL program. We assume that the device integrates $m$ units, and the work items of one workgroup are executed on one unit. Each work item in a workgroup is sequentially executed on one processing element of the unit: altogether $2^{mn}$ work items are executed simultaneously, with one instruction performed in one clock cycle. Since the Abstract OpenCL Platform Model does not care about \textit{warps} (groups of work items for cooperative scheduling in hardware), our Promela model also does not consider warps. Our method can be customized for a particular hardware with its special warp policy by changing the orchestration of processing elements by their unit. Kernel code may contain conditional operators depending on the index of the work item. Without warp scheduling, different branches of such operators are simultaneously executed by the corresponding processing element performing the work item. Work items can use data from both local unit memory and global device memory, but local memory data is not available to the members of other workgroups running on other units.

The key aspect of auto-tuning are the tuning parameters that affect the performance of an OpenCL program. These parameters usually depend on \texttt{size} -- the size of the program's input data; the number of work items also depends on the size of the data. In this paper, we consider the following tuning parameters:
\begin{itemize}
    \item  \texttt{WG} --- the \textit{workgroup size} (defined in the host program). With an optimally selected workgroup size, all processing elements of all units are fully and evenly loaded, which clearly leads to an improvement in the total computation time.
    \item \texttt{TS} --- the \textit{tile size}  (defined in the kernel). The optimal choice of the size of data blocks periodically loaded into the fast local memory minimizes the number of calls to slow global memory, which also leads to faster execution.
\end{itemize}

There exist practical cases when we may have several tile sizes (e.g., when the computation uses matrices of different size), but different sizes of workgroups are usually not required. Our Promela model for computations chooses tuning parameters in a non-deterministic manner by randomly selecting them in the range depending on the input data size. Obviously, the more non-determinism is in the model (in particular with an increase in the number of tuning parameters), the more resources are necessary to perform the model checking. 

Summarizing, our auto-tuning problem is to find the optimal workgroup size \texttt{WG} and tile size \texttt{TS} for the abstract kernel \texttt{abstract\_kernel} (Listing~\ref{kernel}) and its host program, executed on the \texttt{Abstract Platform} as described in Section~\ref{aproc}. For this problem, we now describe the particular steps of our Promela-based approach that relies on finding a counterexample, as outlined in Section~2.

\vspace*{0.5 em}

\noindent \underline{\bf Step 1 of the Counterexample Method}

\vspace*{0.5 em}

The first step is the most time-consuming in our method as it is necessary to take into account all details of executing an abstract parallel OpenCL program on an abstract OpenCL platform. We define the following parallel Promela processes in $Promela$ $Abstract$ $Model$ for modeling this execution:
\begin{itemize}
    \item initial process \texttt{main} selects the values for the tuning parameters and starts the \texttt{host} and \texttt{clock} processes;
    \item process \texttt{host} activates several processes \texttt{device};
    \item every process \texttt{device} launches its subordinate processes \texttt{unit};
    \item every process \texttt{unit} starts processes implementing its processing elements \texttt{pex} and their local memory \texttt{barrier};
    \item every process \texttt{pex} performs computations for the \texttt{abstract\_kernel};
    \item every process \texttt{barrier} locally synchronizes processing elements \texttt{pex};    
    \item service process \texttt{clock} implements global time counting.
\end{itemize}

\begin{figure}[ht!]
	\begin{center}
		\includegraphics[width=0.97\linewidth]{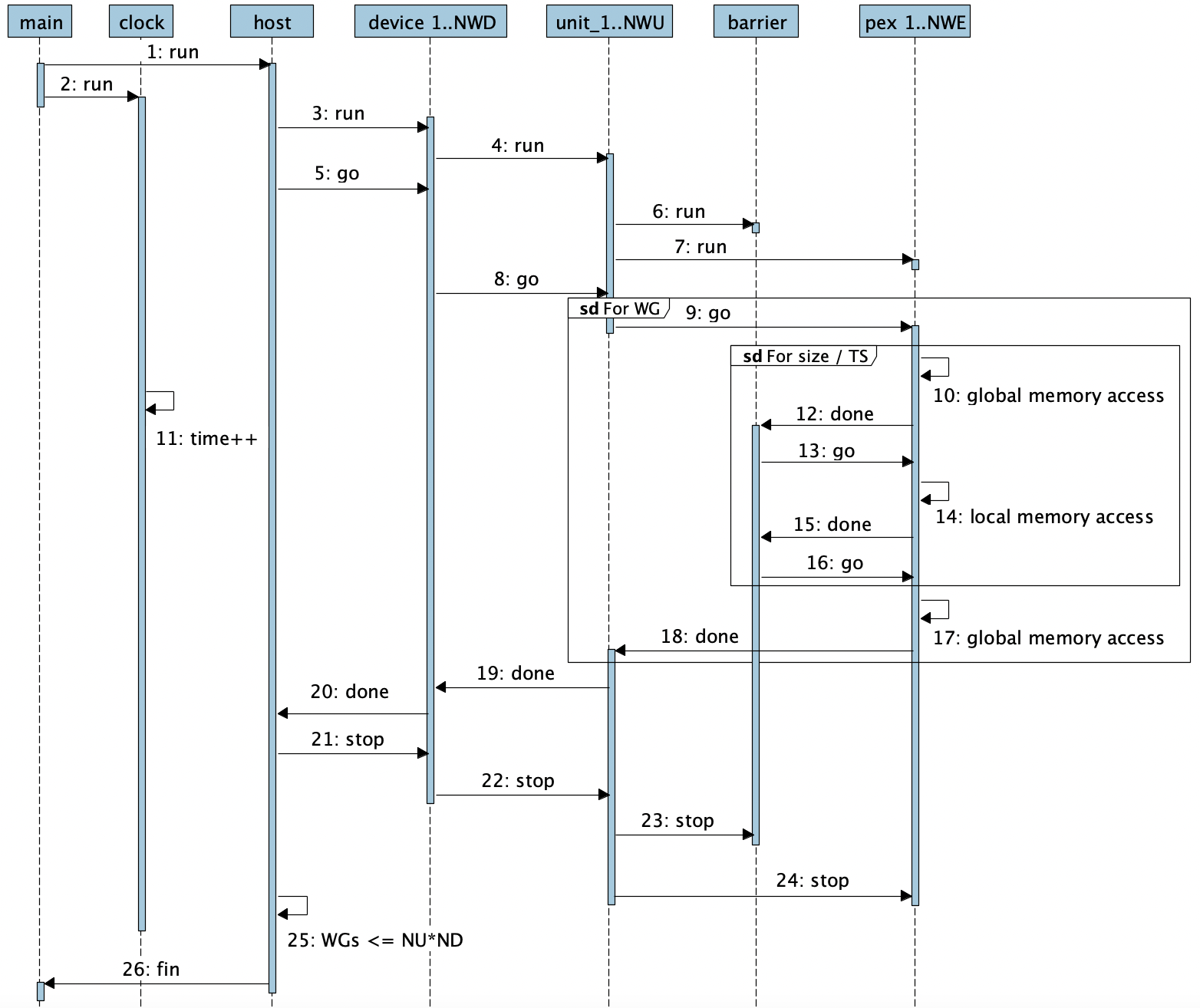}
	\end{center}
	\caption{Sequence diagram for the discussed Promela model}
	\label{msc}
\end{figure} 

Fig. \ref{msc} shows the sequence and communication diagram for these processes. Pairs of processes form master-slave structures: (\texttt{host},  \texttt{device}), (\texttt{device},  \texttt{unit}), (\texttt{unit},  \texttt{pex}), and (\texttt{unit}, \texttt{barrier}). The master starts (\texttt{run}), activates (\texttt{go}) and stops (\texttt{stop}) its slave processes. Slave processes report about termination (\texttt{done}). The processes communicate through handshake channels. In the diagram, we show only one instance for each process. Barrier synchronization is implemented using messages, by waiting for them to be sent by a predefined number of processes. Time is incremented in process \texttt{clock} after the completion of a computation step by all processing elements. Next, we describe our modeling design step-by-step in detail.

In Listing~\ref{main}, we show how process \texttt{main} selects the values for tuning parameters  \texttt{WG} and \texttt{TS} and starts processes \texttt{host} and \texttt{clock}. The number of devices \texttt{ND}, the number of units \texttt{NU} per device, and the number of processing elements \texttt{NP} per unit are global constants declared at the beginning of the model description, as well as factor \texttt{GMT} for the time of computations using the global memory in process \texttt{pex}. In our model, the number of running Promela processes is made up of one main process, one clock, one host, no more than \texttt{ND} devices, no more than \texttt{NU*ND} units, no more than \texttt{NU*ND} barriers, and no more than \texttt{NU*ND*NP} processing elements. We assume for simplicity that data size \texttt{size} is a power of 2. Therefore, the numbers \texttt{WG} and \texttt{TS} are also some random powers of 2. These powers are chosen in lines 14 and 16. Since Promela does not support exponentiation, randomly selected numbers can be obtained by appropriate bitwise shift of \texttt{size}. The number \texttt{NWD} of devices that should be started by the host (lines 14, 16) and the number \texttt{NWU} of units that should be started by every device (line 18) both depend on the number \texttt{WGs} of workgroups. The number \texttt{NWE} of processing elements that should be started by every unit depend on the size \texttt{WG} of a workgroup (line 20). These assignments minimize the number of simultaneously executing processing elements (line 22). In this process, we specify the current value of computation time \texttt{T} (line 23). In line 24, the model of parallel computations starts.

\begin{lstlisting}[label={main},caption={Promela-process for selecting  parameters and launching computation}, style=pml, numbers=left]
active proctype main() { 
 byte i;
 byte n = 10;
  size = 1 << n; // size = 2^n	
  // workgroup size selection
  select (i : 1 .. n-1);
  WG = size >> (n - i);
  // tile size selection
  select (i : 1 .. n-1);
  TS = size >> (n - i);
  // number of working groups
  WGs = size / (WG * TS); 
  // number of working devices 
  NWD = (WGs <= NU * ND -> (WGs / NU) : ND); 	
  // if WGs <= NU 
  NWD = (WGs / NU -> NWD : 1); 				
  // number of working units
  NWU = (WGs <= NU -> WGs : NU); 			
  // number of working elements
  NWE = (WG <= NP -> WG : NP);
  // total number of working elements
  allNWE = NWE * NWU * NWD;
  T = 100;
  atomic {
    run host(); 
    run clock(); 
  } 
}
\end{lstlisting}    

Listing~\ref{host} shows process \texttt{host} that activates \texttt{NWD} processes \texttt{device} in lines 6, 8. If the number of workgroups \texttt{WGs} is relatively small then it waits for their termination (line 11) and marks the termination of the target parallel computations in line 23 by variable \texttt{FIN}. Otherwise, this process reactivates the devices until all workgroups are served (line 17).

\begin{lstlisting}[float, label={host},caption={Promela-process for the host-program}, style=pml, numbers=left]
proctype host () { 
 chan d_hst = [0] of {mtype : action};
 chan hst_d = [0] of {mtype : action};
  FIN = false;
  atomic { 
    for(i : 0..NWD - 1) { run device(d_hst, hst_d); }} 
  atomic {
    for(i : 0..NWD - 1) { hst_d ! go; }}
  if
  :: WGs <= (NU * ND) -> atomic { 
          for(i : 0..NWD - 1) { d_hst ? done; hst_d ! stop; }}
  :: else -> 
          for(i : 0..WGs / (NU * ND) - ND) {
                atomic { 
                d_hst ? done;
                allNWE = allNWE + NWE * NWU;
                hst_d ! go; }}
          for(i : 0..ND - 1) { 
                atomic { 
                d_hst ? done;
                hst_d ! stop; }}
  fi
  FIN = true;
}
\end{lstlisting}  

As an extension of our previous paper \cite{garanina2022model}, we describe here in detail processes \texttt{device} and \texttt{unit}. \textit{Process} \texttt{device} (Listing \ref{muproc}) starts its \texttt{NWU} instances of process \texttt{unit} in line 4, activate them in line 7 and awaits their termination in lines 11, 15, or 18, depending on the number of workgroups, like the unit process above waits their \texttt{pex}es. If the slave unit does not suppose to be reactivated, then the number of simultaneously working processing elements is decreased (lines 11 and 18). The device also reports about termination to its master -- the host process (line 20). After stopping, the device stops all its units (line 22).
\begin{lstlisting}[float, label={muproc},caption={Promela-process for a device}, style=pml, numbers=left]
proctype device (chan d_hst; chan hst_d) { 
 chan dev_u = [0] of {mtype : action};
 chan u_dev = [0] of {mtype : action};
  atomic{ for(i : 0..NWU - 1) { run unit(dev_u, u_dev); }} 
  do 
  :: hst_d ? go ->
        atomic{ for(i : 0..NWU - 1){ dev_u ! go; }}
        if 
        :: WGs <= NU -> 
              atomic { for(i : 0..NWU - 1) {
                          u_dev ? done; allNWE = allNWE - NWE; }}
        :: else -> 
              atomic { 
                 for(i : 0..WGs - 1) {
                     u_dev ? done; dev_u ! go; }}
              atomic { 
                 for(i : 0..NU - 1) {
                     u_dev ? done; allNWE = allNWE - NWE; }}
        fi
        d_hst ! done;
  :: hst_d ? stop -> 
        atomic{ for(i : 0..NWU - 1) { dev_u ! stop; }} 
        break;
  od
}	
\end{lstlisting}    

\textit{The process} \texttt{unit} (Listing \ref{unite}) starts  its \texttt{NWE} processing elements \texttt{pex} and their local memory \texttt{barrier} (lines 6-8). Then upon receiving the activating signal from its master device, the unit process activate its \texttt{pex}es (line 11) and then waits termination of their work (line 14). If the size of every work group \texttt{WG} is less then the number \texttt{NP} of slave processing elements of the unit, after receiving reports from all activated \texttt{pex}es, the process \texttt{unit} goes to report about the termination to its master device in line 23. If it is not the case, the unit again activates the just finishing processing elements in line 17 until all all work items in the workgroup have been completed. Upon receiving the stop signal from its device, the unit process stops its barrier and all processing elements (lines 25-26).

\begin{lstlisting}[label={unite},caption={Promela-process \texttt{unit} for abstract unit}, style=pml, numbers=left]
proctype unit (chan dev_u; chan u_dev) { 
 chan pex_b = [0] of {mtype : action};
 chan b_pex = [0] of {mtype : action};
 chan pex_u = [0] of {mtype : action}; 
 chan u_pex = [0] of {mtype : action}; 
  run barrier (pex_b, b_pex);
  atomic { for (i : 0..NWE - 1) {
               run pex(i, pex_b, b_pex, pex_u, u_pex); }}
  do 
  :: dev_u ? go ->
          atomic { for(i : 0..NWE - 1) { u_pex ! go; }}
          for(i : 0..WG - 1) {
                 atomic { 
                   pex_u ? done;
                   if
                   :: WG <= NP -> skip;
                   :: else -> u_pex ! go;
                   fi }}
          if
          :: WG <= NP -> skip;
          :: else -> atomic { for(i : 0..NP - 1) { pex_u ? done; }}
          fi
          u_dev ! done;
  :: dev_u ? stop -> 
          pex_b ! stop; 
          atomic { for (i : 0..NWE - 1) { u_pex ! stop; }}
          break;
  od
}
\end{lstlisting} 
 
Listing \ref{barrier} shows the \textit{local synchronization process} \texttt{barrier}: it synchronizes processing elements of one unit using the \texttt{pex\_b} and \texttt{b\_pex} channels. It receives a message from the \texttt{pex} processes about the suspension of their work and counts in the \texttt{i} variable the number of processes that are waiting for the termination of processing the local memory of other members of the group currently performing calculations (line 4). When it is equal to the number of processing elements \texttt{NWE} launched by its master unit, the barrier is considered passed and in line 5 the the barrier process allows the processing elements to continue computations.

\begin{lstlisting}[label={barrier},caption={Promela-process \texttt{barrier} for local synchronization}, style=pml,  numbers=left]
proctype barrier (chan pex_b; chan b_pex) { 
  do 
  :: pex_b ? done ->
           for(i : 1..NWE - 1) { pex_b ? done; } 
           atomic { for(i : 0..NWE - 1) { b_pex ! go; }} 
  :: pex_b ? stop -> break;
  od
}
\end{lstlisting}  

Listing~\ref{pex} shows in detail the work item process \texttt{pex} that performs computations of an instance of the kernel code \texttt{abstract\_kernel} presented in Listing~\ref{kernel}. The total number of these processes that are created in the Promela model depends on the size of the input data. The number of simultaneous \texttt{pex} processes depends on the number of units, which is significantly smaller than the total number of processes. Each of them is started by its own unit, whose Promela process is described in \cite{git}. 
\begin{lstlisting}[label={pex},caption={Promela-process \texttt{pex} for a processing element}, style=pml,  numbers=left]
inline long_work (gt, tz) {
  do  
  :: time > (start_time + (gt * tz)) -> break;
  :: else -> atomic { 
               cur_time = time;
               NRP_work++;
               time == cur_time + 1;} //wait
  od 
}
proctype pex (byte me; chan pex_b; chan b_pex; 
                    chan pex_u; chan u_pex) { 
   do 
   :: u_pex ? go ->
      atomic { start_time = time; cur_time = time; }
      for(i : 0..(size / TS - 1)) { 
          long_work(GMT, TS) // access to global memory
          pex_b ! done; b_pex ? go; // waiting for local co-workers  
          start_time = time;
          if  // 'if' access to local memory
          :: me % 2 -> long_work(1, TS) 
              // 'else' access to local memory
          :: else -> long_work(1, TS) 
          fi
          pex_b ! done; b_pex ? go; // waiting for local co-workers
      }
      start_time = time;
      long_work(GMT, 1) // copy the result to global memory
      pex_u ! done;  
   :: u_pex ? stop -> break;
   od
}
\end{lstlisting}    

Process \texttt{pex} is connected by the handshake synchronization channels \texttt{pex\_b} and \texttt{b\_pex} with the local group barrier, and \texttt{pex\_u} and \texttt{u\_pex} with the master unit. In these channels, the work item receives start and stop commands, and also informs about the termination of computations. In our model, we abstract from the specific computations that the kernel performs, so we only use the computation time. We also assume that a computation in local memory takes one time unit, and a computation using global memory takes \texttt{GMT} units of time. This abstraction can be straightforwardly refined for a particular program and hardware.

Thus, in process \texttt{pex} we model only the number of computational steps performed by the kernel (the \texttt{for} loop in line 15), depending on the size of the input data and calls to global memory (lines 16 and 27) and to local memory (lines 19-23). Upon termination of the computation step, \texttt{pex} reports this event to the process that implements global time by increasing the counter of the currently running \texttt{NRP\_work} processes (line 6 of inline macro \texttt{long\_work}). Note that due to the blocking semantics of the Promela language, the process can proceed to the next stage of its computation only when the global time \texttt{time} is increased by 1 (line 7 of \texttt{long\_work}). The ``long work'' modeling of  computations of the abstract kernel function \texttt{f}, \texttt{g1}, \texttt{g2} and \texttt{h} finishes after \texttt{gt*tz} time units, depending on the type of memory accessed. Synchronization on the local barrier occurs twice, as in the original kernel: line 10 of the kernel (Listing~\ref{kernel}) corresponds to line 17 of the model (Listing~\ref{pex}), and line 17 corresponds to line 24. Process \texttt{pex} terminates (line 28) after writing its result into global memory.

Service \textit{clock process} \texttt{clock} implements global time counting. This process increments the global \texttt{time} counter variable when all running \texttt{pex} processes (their number \texttt{allNWE}) reports by increasing the \texttt{NRP\_work} shared variable that currently they are computing. The \texttt{clock} process stops when the global variable \texttt{FIN} is \texttt{true}. The \texttt{time} value at this moment is the time taken for the computation.

\begin{lstlisting}[label={clock},caption={Promela-process for global time}, style=pml, numbers=left]
proctype clock () { 
  do
  :: FIN -> break;
  :: allNWE != 0 && NRP_work == allNWE -> 
            atomic { NRP_work = 0; time++; }
  od
}
\end{lstlisting}

\noindent \underline{Step 2 of the Counterexample Method}

In specializing the over-time property $\Phi_o$ for the auto-tuning problem, we use the value of variable \texttt{FIN} that marks the end of calculations, and the final value \texttt{time}. The over-time formula without superindex is used in the model checking approach to general auto-tuning problem, while this formula with superindex ‘$p$’ is specialized for our method which uses Promela model. The SPIN property specification language is  temporal logic LTL, such that we can write: $$\Phi_o^p = \textbf{G} (\texttt{FIN} \rightarrow (\texttt{time} > T)),$$which corresponds to the statement ``Always when the parallel program terminates, its execution time is greater than $T$.''
\vspace*{0.5 em}

\smallskip

\noindent \underline{Step 3 of the Counterexample Method}

\smallskip

The third step of our method finds the minimal time for which a parallel computation program can terminate. This step begins with launching the SPIN verifier with the constructed Promela Abstract Model and the formula $\Phi_o^p$ for some value of time $T$. We then decrease the value of $T$ until the SPIN stops generating counterexamples, i.e., until it agrees that the program cannot be executed in a time shorter than the final $T$. In our approach, we find the execution path with minimal model time. This path may not be the shortest one due to many model steps which do not increase the value of variable \texttt{time}. We indeed came across such paths in our experiments. The initial value of $T$ can be found using the simulation mode in SPIN. During simulation, SPIN reproduces one of the finite scenarios of the system behaviour, fixing the values of the variables used in the model at the end of the simulation. Therefore, we can use value \texttt{time} which corresponds to the end of the program in the simulated scenario. To decrease the value of $T$ in the next SPIN runs, we may use the bisection method shown in Fig.~\ref{Bisect}. The final counterexample gives the minimal model time value.
\vspace*{0.5 em}

\pagebreak

\noindent \underline{Step 4 of the Counterexample Method}

The final step of our approach is to analyze the last counterexample to extract the optimal configuration of tuning parameters. For counterexample analysis, SPIN provides running a simulation corresponding to the counterexample. In the auto-tuning problem, there is no need to search for the optimal computation path. Therefore, there is not necessary to analyze the transitions of the final counterexample. To solve our problem, we only need to extract the values of the tuning parameters \texttt{WG} and \texttt{TS}, which are known in the final counterexample simulation. The final values of the tuning parameters \texttt{WG} and \texttt{TS} form the optimal parameter configuration we are looking for.

\section{Adaptation for Limited Computation Resources: Using Swarms}\label{limited}

As our solution requires a large number of interactions through channels, its straightforward implementation does not scale well: with an increase in the size of the input data, the memory required for the verification process in Step 3 begins to grow and may exceed the available physical memory of the computer, which we sometimes observed in our experiments. 

We address the problem of limited resources by using \textit{the swarm model checking} method proposed in \cite{holzmann2008tackling,holzmann2010swarm}. This method uses state hashes and therefore does not visit all possible states of the system. The swarm model checking starts several different verification tests checking the specified property during bounded time and with a bounded number of model steps. Similar to the standard model checking method, it generates counterexamples when one of the paths violates the property. The search of such a path can be parallelized across processor cores and even performed on different network nodes. In \cite{staroletov2020model}, we already applied this approach to the automatic solution of various puzzles and practical problems. 

In order to use swarm model checking for alternative non-bisection searching, we can formulate \textit{the non-termination property $\Phi_t$} for the impossibility of program termination as follows: ``A parallel program \textit{cannot} terminate'' (LTL formula $\Phi_t = \textbf{G} (\neg\texttt{FIN})$). We definitely know that our parallel programs that compute some output data must terminate. Hence, we expect that there is at least one counterexample which demonstrates that the program terminates and provides the termination model time. Swarm model checking produces several  termination model times from several counterexamples, from which we can choose the minimal termination model time. To be sure that there is no smaller time, we launch successively swarm model checking with a decreasing model time and over-time property $\Phi_o^p$ in the following manner: we stop searching for the minimal termination model time if swarming with the current model time provides no results within the previous swarm execution time or this swarming finds no smaller times. In other words, the criterion for stopping the search for the optimal time is the ability of the SPIN swarm to find counterexamples, rather than the number of such findings. If the swarm does not find a counterexample as quickly as at the previous swarm launching, the counterexample with a smaller \texttt{time} value does not exist with very high probability. In future work, we plan to formally prove that the configuration of tuning parameters found by the swarm is indeed optimal.

\pagebreak

\begin{figure}[ht]
	\begin{center}
	\includegraphics[width=0.8\linewidth]{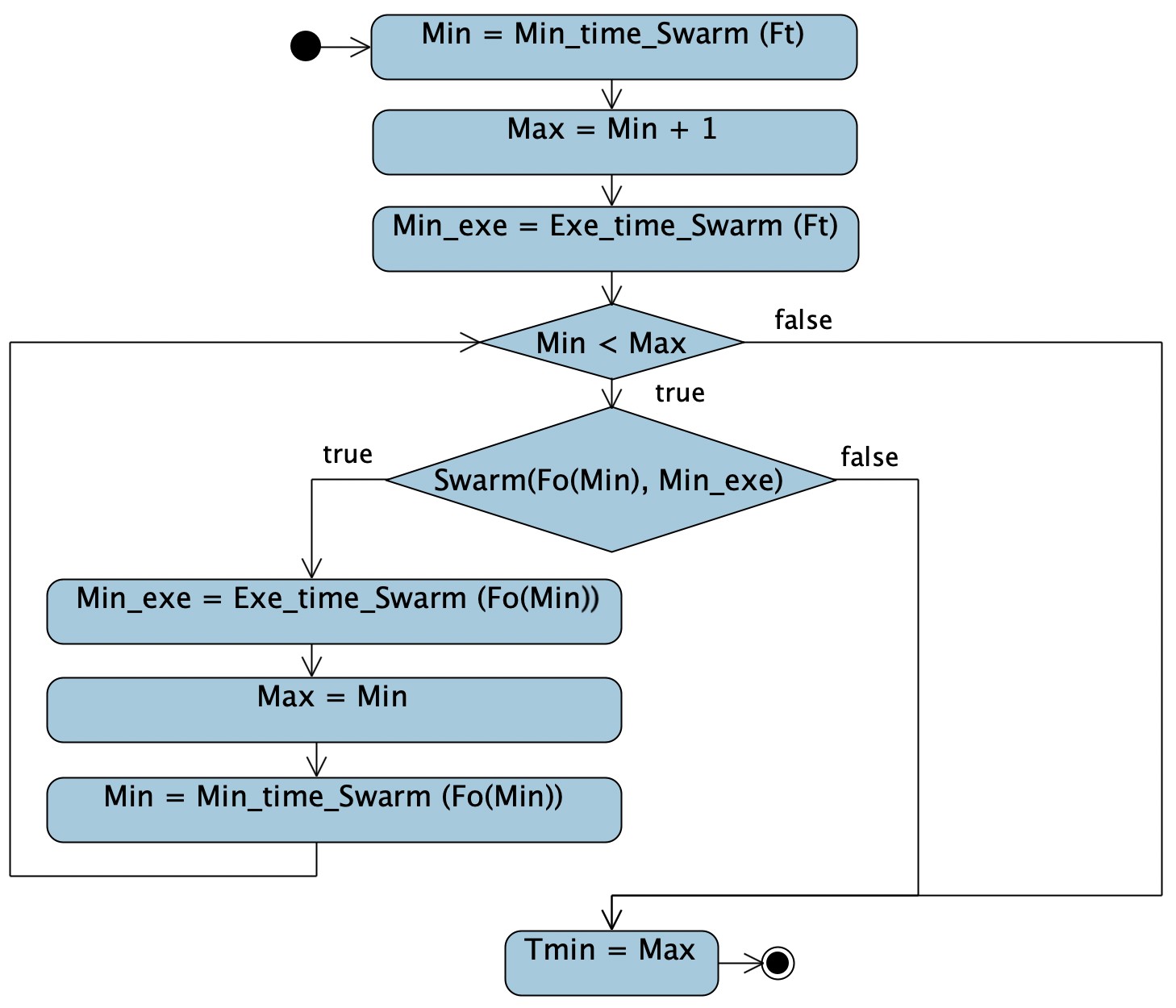}
	\end{center}
	\caption{The swarm search method}\label{SwarmSeach}
\end{figure}

Our algorithm in Figure~\ref{SwarmSeach} implements the described search strategy with two functions: \texttt{Min\_time\_Swarm(F)} and \texttt{Exe\_time\_Swarm(F)} for the minimal model time found by the swarm for formula \texttt{F} and the execution time of this swarm, respectively. Predicate \texttt{Swarm(F,exe\_time)} is true iff the swarming formula \texttt{F} finishes within \texttt{exe\_time}. Formula \texttt{F\_t} is for the non-termination property, and formula \texttt{F\_o(T)} is for the over-time property with time \texttt{T}.

In our implementation, we have developed a script that simulates each counterexample found by the swarm and finds values of \texttt{time}, \texttt{TS} and \texttt{WG} in the output. We also provide a simple program which, after the script has finished, sorts these counterexample results by \texttt{time} values with corresponding tuning parameters \texttt{TS} and \texttt{WG}. 

We make an assumption (which corresponds to the situation in practice) that every device and every unit of a device work in exactly the same manner, i.e., it takes exactly the same time (and other resources) to perform computation of the same complexity. Hence, to estimate the computation time, we can abstract from the number of devices and the number of units in these devices and consider just one device and one unit with all its processing elements. Note that we cannot consider just one processing element because, in general, they compute different functions of different complexity depending on their local identifiers while the devices and units synchronize their work independently of the computation complexity. With this abstraction, the number of Promela processes is made up of one main process, one clock, one host, one device, one unit, one barrier and several processing elements. In the model used in our experiments \cite{git}, the total number of Promela processes is 10. Here, the only really large parameter is the input data size.

\section{Experiments with the Promela Abstract Model}\label{exprm}

In Table \ref{table1} we show the results of a series of experiments conducted with the SPIN model checker (version 6.5.1). During the tests, we studied different sizes of input data for a platform architecture with the following parameters: one device, one unit and four processing elements. 

\begin{table*}[htbp]
\caption{Resulting parameter values and memory consumption measured in the experiments}\label{table1}
\begin{center}
\begin{tabular}{ |p{0.1cm}|p{0.46cm}|p{0.9cm}|p{1.3cm}|p{0.4cm}|p{0.4cm}|p{1.5cm}|p{1.3cm}|p{0.7cm}|p{0.6cm}|p{0.8cm}|}
\hline
\textbf{N}&{\textbf{Size}}&{\textbf{Model time}}&{\textbf{Steps}}&{\textbf{TS}}&{\textbf{WG}}&{\textbf{Memory usage (exhaustive mode)}}&{\textbf{Memory usage (swarm mode)}}&{\textbf{Verifi-ca-tion time}}&{\textbf{1st trail in}}&{\textbf{1st trail optimality}} \\
\hline
1 &	8 &	44 & 1658 & 4 & 4 & 0.280GB & - & 2s  & 1s & 84\%
\\
\hline
2 &	16 & 156 & 5673 & 4 &	8 &	2.081GB & - & 25s & 1s & 65\%
\\
\hline
3 &	32 & 584 & 21011 & 4 & 16 &	14.767GB & 115MB & 25m/ 1h & 7s/ 1s & 79\%/ 77\%
\\
\hline
4 &	64 &	2224 & 71495 &	8 &	32 &	- &	128MB & 1h & 1s & 83\%
\\
\hline
5 &	128 &	9344 & 267119 &	64 &	64 &	- &	172MB & 1h & 1s & 94\% 
\\
\hline
6 &	256	& 36234 & 1300634 &	4 &	4 &	- &	1.145GB & 2h & 3s & 94\%
\\
\hline
7 &	512	& 142090 & 5099397 & 4 & 4 & - & 2.367GB & 2h & 3s & 99\%
\\
\hline
8 &	1024 &	549912 & 15973533 &	32 & 16 & - & 	16GB & 4h & 3m & 99\%
\\
\hline
\end{tabular}
\end{center}
\end{table*}

We observe in the table of results that, up to input data \texttt{size} 32, exhaustive verification with over-time formula $\Phi_o^p$ was possible for our machine (an M1-based MacBook with 16Gb of RAM).

This auto-tuning finds optimal values of parameters with the shortest simulation time (parameter which is also calculated in our model). For reference, we include model steps, shown in SPIN simulation mode for the best counterexample. When the data size is significantly larger, we may hit the 16 GB memory threshold. Therefore, we run the swarm verification with non-termination formula $\Phi_t$ on 1-8 cores for given maximum verification time threshold, and the swarm process is now able to find a large number of counterexamples with trails to the final state \texttt{FIN} for larger data sizes. We also developed a runner script to process all the trails found by SPIN. It simulates a particular trail, obtains the simulation time and TS/WG parameters from the simulation output, sorts them by time and required steps. Using such a script  finally helps to obtain the minimal time and optimal parameters from a bunch of trails. Getting multiple trails is possible due to \texttt{-e} parameter (create trails for all errors) during the verification and swarm run. We also had to increase maximum search depth depending on the input size with the last one set to $2\times 10^8$ by option \texttt{-m}. 

In Table~\ref{table1}, ``Model time'' is the time that is calculated in the Promela model (variable \texttt{time}), and ``Verification time'' is the time spent by the SPIN verifier or its swarm processes. In the last two columns of the table, we show the time required to obtain the first counterexample with a trail -- the quickest sub-optimal solution of the problem (both using exhaustive and swarm methods) and its optimality as the ratio of the optimal solution modeling time to the time of the first counterexample (if the times coincide, we also take into account the number of steps spent).

Therefore, we can conclude that our method is feasible in practice: for small dimensions, we can prove that our interacting system terminates in the correct order. Once we hit the memory limits in the model checker, we can use the swarm method and get the approximate model time and target parameters  with a full trace of execution after a series of quick model runs. Of course, this method is not yet applicable for modeling operations with really big data. However, we initially set ourselves goals to consider modeling with an abstraction from operations and data, since we are primarily interested in the sequence of their processing.

\section{Application Use Case: Auto-Tuning the Minimum Problem}\label{apps}

In this section, we demonstrate how our approach to auto-tuning based on model checking works for a particular application use case.
Our use case is the problem of computing the minimal value in a very big integer array, we call it the \textit{Minimum problem}.

We offer an OpenCL kernel program for the Minimum problem, run and tune parameters for it using Nvidia GPU P104-100 of the Pascal architecture.  Then we show how the general Promela model defined in Section \ref{spin} can be modified to model a particular case of this parallel computation problem. Ultimately, we apply our method of counterexamples and thereby find the optimal tuning parameters for the Minimum problem.

\subsection{OpenCL Program for the Minimum Problem}

The Minimum problem belongs to the practically relevant class of computing problems which compute a single result for an input array; they are also known as reductions. Such problems are popular in teaching parallel programming of parallel programs for massively parallel message systems like MPI. On the other hand, they are also well suitable for parallel computing on massively parallel systems like GPU. The usual approach is to first scatter parts of the array over compute units, which then further divide them and compute the minimum in their part, after which it is necessary to collect all parts and calculate the global minimum.

   \begin{figure}[ht]
	\begin{center}
	\includegraphics[ width = 0.8\linewidth]{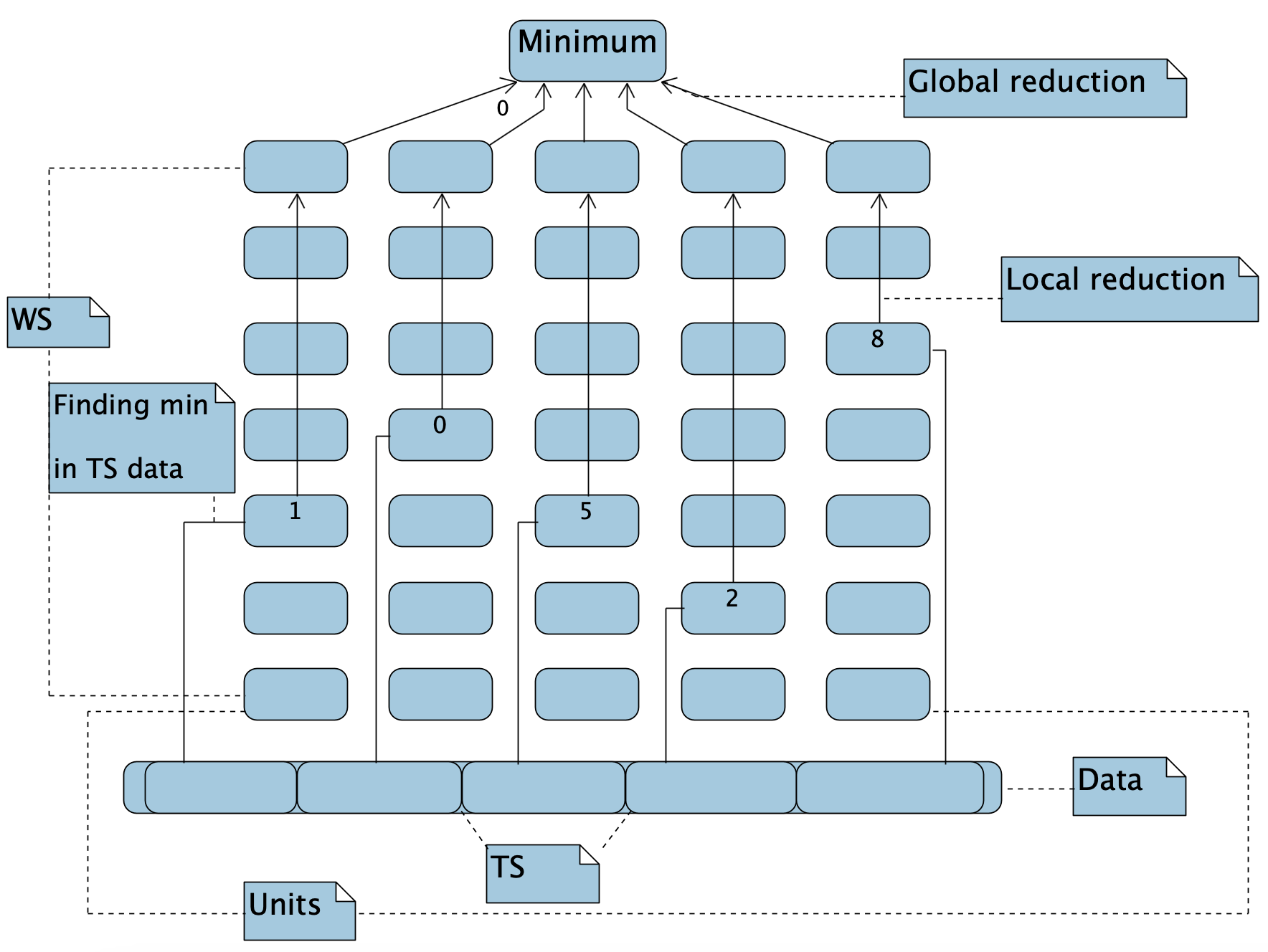}
	\end{center}
	\caption{The propagation of the minimum in our solution}\label{propog}
\end{figure}

The solution scheme is shown in Figure~\ref{propog} and presented in Listing \ref{KernMin} for the kernel and Listing \ref{HostMin} for the host. In our implementation, we work on a grid of the total number of processing elements and this number is set from the host program as $global\_item\_size$ (see host, line 10). The current work item number is available from the kernel as $get\_global\_id(0)$. The total number of processing elements is divisible into workgroups each by $get\_local\_size(0)$, which is referred to $WG$ in our programs. We can consider all processing elements as a two-dimensional array with coordinates available as $(get\_global\_id(0) / get\_local\_size(0), get\_local\_size(0))$ (see kernel, lines 3 and 4). At the same time, according to the first coordinate, we can assume that some abstraction on smart computing units is done here, where each of them has local memory and $WS$ child processing elements during computations.

In general, each of the processing element receives a portion of data of size $TS$, while it can be set relative to the splitting of all data into equal parts or relative to the number of such splits and then recalculated into array elements. Then the parallel algorithm finds all minima for each calculator in its portion of data of size $TS$ (kernel, lines 7-9) and saves them in local memory. Further, all found minima are compared started from zero cells of local memory (kernel, lines 12-16) and then unloaded into global memory, where the final reduction is carried out for simplicity on the host (host, lines 19-24). The full program for the minimum problem is available in \cite{git}.

\pagebreak

\begin{lstlisting}[label={KernMin},caption={The OpenCL kernel for the Minimum problem}, style=cplusplus, numbers=left]
__kernel void findMinValue(__global int* glob, __global int* mins, __local int* loc, int TS) {
#define MIN(a, b) if (a > b) a = b;
  int my_unit = get_global_id(0) / get_local_size(0);
  int my_elem = get_local_id(0);
  loc[my_elem] = INFINITY;
  // MAP
  for (int i = 0; i < TS; i++) {
    MIN(loc[my_elem], glob[i + get_global_id(0) * TS])
  }
  barrier (CLK_LOCAL_MEM_FENCE);
  // REDUCE local
  if (my_elem == 0) {
    mins[my_unit] = loc[0];
    for (int i = 1; i < get_local_size(0); i++) {
      MIN(mins[my_unit], loc[i])  
    }
  }
}
\end{lstlisting}

\begin{lstlisting}[label={HostMin},caption={The host code for the Minimum problem}, style=cplusplus, numbers=left]
int main(...) {
  ...
  size_t units = ...; // Should be near the value of SM of the card
  size_t global_item_size = units * WG; //Total number of calcs
  size_t local_item_size = WG; //Work group size
  size_t TS = ...; //Tile size
  size_t data_size = ...;// clArray[data_size] is our Big Data
  ...
  clSetKernelArg(kernel, 0, sizeof(cl_mem), &clArray);
  clSetKernelArg(kernel, 1, sizeof(cl_mem), &clMins);
  clSetKernelArg(kernel, 2, 
                        sizeof(cl_uint) * local_item_size, NULL);
  //Run the kernel on the grid (global_item_size x local_item_size)  
  clEnqueueNDRangeKernel(queue, kernel, 1, NULL, &global_item_size,
                                &local_item_size, 0, NULL, NULL);
  TS = data_size / (units * WG); //Chunk of data per TS
  clSetKernelArg(kernel, 3, sizeof(cl_uint), &TS);
  ...						   
  clEnqueueReadBuffer(queue, clMins, CL_TRUE, 0, units *  
                            sizeof(cl_uint), mins, 0, NULL, NULL);
  int min = INT32_MAX;
  // REDUCE global
  for (int i = 0; i < units; i++) {
    if (mins[i] < min) min = mins[i]; //Total minimum = min
  }
}
\end{lstlisting}

Let us now discuss the choice of optimal parameters for running the kernel on particular hardware. We are going to conduct a series of experiments on the GPU P104-100 to find the minimal element in a 4GB array using the proposed kernel, while changing the values of the tuning parameters in nested loops. The key features of our target platform are described in the white paper on the Pascal architecture \cite{pas}, and its current configuration is presented, for example, in \cite{vid}. 

Our target GPU has a total of 1920 processing elements (or CUDA cores according to Nvidia terminology). Respectively, the total number of elements in the grid of the OpenCL program (how many instances of our kernel will be launched) should be commensurate with this value.  We should also pay attention to the number of streaming multiprocessors (SM), which correspond to the compute units in the OpenCL platform; here there are 15 of them. Analyzing their structure, we can see that in general, inside each SM there are 128 processing elements for different purposes, which actually form the given configuration of CUDA cores (1920 = 128 * 15). In addition, there is also 48Kb L1 local memory per each SM. From this we can conclude that in order to effectively execute our kernel on our hardware platform, we need to set the following initial configuration: units = 15; WG = local\_size = 128, global\_size = 15 * 128. However, it should be noted that this launch configuration is only the initial one: some processing elements may remain idle while waiting for data. The task of the OpenCL compiler and the runtime environment is to efficiently use all available resources for a given kernel and launch configuration, and our task is to select such parameters as a result of several launches, auto-tuning or exemplary models, such as ours in Promela.

\subsection{From the Minimum OpenCL Kernel to its Promela Model}
In this section, we describe how we can obtain a Promela model for the minimum problem, as the first step of our auto-tuning approach based on model checking with counterexamples. We should adapt all Promela processes described in Section \ref{spin} except \texttt{clock} and \texttt{host} processes, as follows below. The full Promela model for the minimum problem is available in \cite{git}.

First, we consider the service \texttt{main} process. Before selecting optimization parameters (from line 3, Listing \ref{main}), the process loads data to global memory and updates the local memory of all units with maximal values \texttt{MAX} of data type. Arrays \texttt{glob} and \texttt{loc} are defined as global Promela variables.

\begin{lstlisting}[label={mainMin},caption={A part of \texttt{main} Promela-process in the Minimal model for defining initial data}, style=pml,numbers=left]
active proctype main() { 
...	
  // loading input data to global memory
     for(i : 0 .. size-1) { glob[i] = size - i; }
  // preparing local memory 
     for(i : 0 .. LM*NU-1) { loc[i] = MAX; }
  // workgroup size selection
...
}
\end{lstlisting}

The \texttt{host} process in case of the Promela model for the minimum problem uses only one device and does not operate on global memory, so we do not have to change it. There is also nothing to change in the service \texttt{clock} process.

The \texttt{device} process must count the number \texttt{nwg} of each workgroup launched on one compute unit and send it to its compute units in order to correctly indexing global memory in processing elements. Hence, we change the type of channel \texttt{dev\_u}: instead of just saying 'go' to units (line 11 in Listing \ref{muproc}), a device computes and sends information about the number of  workgroups in lines 9-10, Listing \ref{muprocMin}.

\begin{lstlisting}[label={muprocMin},caption={A part of \texttt{device} Promela-process in the Minimal model for sending workgroup numbers}, style=pml,numbers=left]
proctype device (chan d_hst; chan hst_d) { 
 chan dev_u = [0] of {byte, mtype : action};
 ...
 for(i : 0..WGs-1) {
      atomic { u_dev ? done; 
               dev_u ! nwg, go;
               num++;
               if 
               :: num >= NU -> nwg++; num = 0;
               :: else -> skip;
               fi }}
 ...
}	
\end{lstlisting} 

For correct indexing of local memory in the Promela model, the \texttt{unit} process must inform its processing elements about its number \texttt{mu} provided by its device. So we change the corresponding \texttt{proctype} definitions and \texttt{run} commands (line 6, Listing \ref{unitMin}).
Beside the number of a workgroup received from its device (line 8, Listing \ref{unitMin}), the unit must send to each its processing element information about the number \texttt{iter} of launched work items assigned to it within one group computation for correct global indexing (lines 9 and 20, Listing \ref{unitMin}). We correspondingly change a type of channel \texttt{u\_pex} (line 3, Listing \ref{unitMin}). If a processing element does not finish compute assigned work items, the unit launch it again (line 20, Listing \ref{unitMin}). After finishing, all processing elements report about it in line 32, Listing \ref{unitMin}.

\begin{lstlisting}[label={unitMin},caption={Promela-process \texttt{unit} in the Minimal model}, style=pml, numbers=left]
proctype unit (byte me, chan dev_u; chan u_dev) { 
 ...
 chan u_pex = [0] of {byte, mtype : action}; 
 run barrier (pex_b, b_pex);
 atomic { for(i : 0..NWE-1) { 
               run pex(i, me, pex_b, b_pex, pex_u, u_pex); }}
 do 
 :: dev_u ? nwg, go ->
  atomic { for(i : 0..NWE-1) { u_pex ! nwg, gowg; u_pex ! 0, go; }}
    if 
    :: WG <= NP -> atomic { for (i : 0..NWE-1) { pex_u ? ewg; }}
    :: else -> 
            iter = 1;
            num = 0;
            for(i : 0..WG-NP-1) { 
                  atomic { 
                    pex_u ? status;
                    if 
                    :: status == end_wg -> skip;
                    :: else -> u_pex ! iter, go;
                    fi
                    num++;
                    if 
                    :: num >= NP -> iter++; num = 0;
                    :: else -> skip;
                    fi	}}
            for(i : 0..NWE-1) { pex_u ? ewg; }
    fi	
    u_dev ! done;
 :: dev_u ? 0, stop -> 
    atomic { for(i : 0..NWE-1) { u_pex ! 0, stop; }}
    atomic { for(i : 0..NWE-1) { pex_u ? done; }}
    pex_b ! stop; 
    break;
 od
}
\end{lstlisting}  

There is exactly one local memory barrier at the end of computation of every unit. This barrier works for one processing element only (the first element of the unit), so, in line 5 of \texttt{barrier} process (Listing \ref{barrier}) we replace the collective launching by individual launching by removing the \texttt{for} loop.

The processing element \texttt{pex} uses local information about its local id and its unit id from its \texttt{proctype} definition and information received from its unit about a current workgroup number \texttt{nwg} and iteration \texttt{iter} (lines 6, 8, Listing \ref{pexMin}) in order to index and compare local and global memory data and save the minimum in the corresponding slot of the local memory (line 15, Listing \ref{pexMin}). Comparing with abstract \texttt{pex}, at this stage of the minimum problem the process does not care about local computations of its co-workers, so it does not need a barrier. Every processing element \texttt{pex} must inform its unit about the end of processing assigned work items by sending the message in line 18, Listing \ref{pexMin}.
When the processing element with zero local id have received from its master unit information about the end of computation (line 9, Listing \ref{pexMin}), it reduces the collective result from the  local memory (lines 14-16, Listing \ref{pexMin}) and saves it to the first slot of global memory (line 17, Listing \ref{pexMin}). 

\begin{lstlisting}[label={pexMin},caption={Promela-process \texttt{pex} in the Minimal model}, style=pml, numbers=left]
proctype pex (byte me; byte mu; chan pex_b; chan b_pex; chan pex_u; chan u_pex) { 
...
byte myloc = me + mu*NP;      // my place in local memory

do 
:: u_pex ? nwg, gowg ->
   do 
   :: u_pex ? iter, go ->
      atomic { start_time = time;
               cur_time = time; }
      glob_id = (WG > NP -> nwg*WG + me + iter*NP : nwg*WG + me);
      shift = glob_id * TS;
      // MAP
      for(i : 0 .. TS-1) {
            min(loc[myloc], glob[i + shift])
            long_work(GMT) } // access to global memory
      if 
      :: (iter+1) >= WG/NP -> pex_u ! end_wg; break;
      :: else -> pex_u ! n_end_wg;
      fi
    od
:: u_pex ? 0, stop -> 
   pex_b ! done;
   if 
   :: me == 0 ->
      b_pex ? go;             // waiting for local co-workers 
   // REDUCE local
      for(i : 1..(NWE-1)) { 
            min(loc[myloc], loc[myloc + i]) 
            time++;	}         // access to local memory
   // copy the result of this working group to global memory
      min(glob[0], loc[myloc]) 
      time = time + GMT;
   :: else -> skip;
   fi
   pex_u ! done;
   break;
od
}
\end{lstlisting}  

We can obviously use the outlined adaptation technique for a broad variety of computing problems based on reduction, with different base operations (which in our use case is the binary \texttt{min} operation).

\subsection{The experimental results for the Minimum Problem}

In this section, we present the experimental results for two possible ways of finding optimal configuration of tuning parameters for the use case of Minimum problem. The first way is finding the optimal parameter values manually by running the parallel implementation on the particular architecture (in our case --- the Nvidia P104-100 GPU) for different combinations of parameter values and comparing the run times, i.e., this is the way still often used by application programmers in practice. The second way of finding the optimal configuration of the tuning parameters is applying our approach presented in this paper: using the Promela model of the OpenCL code and looking for a counterexample by means of the model checking tool SPIN. 

In the manual approach, a series of tests were conducted to find the minimum for the same data size, but with different kernel startup parameters. The time and speed of processing data of a given length were considered when the \texttt{clEnqueueNDRangeKernel} (kernel launch) and \texttt{clEnqueueReadBuffer} (getting an array with minima for reduction on the host) functions worked on previously prepared data. 

\begin{table*}[htbp]
\caption{Experiments with the Minimum kernel}\label{table2}
\begin{center}
\begin{tabular}{ |p{1cm}|p{1cm}|p{1cm}|p{1cm}|p{1cm}|p{2cm}|}
\hline
\textbf{N}&{\textbf{Global size}}&{\textbf{WG}}&{\textbf{TS}}&{\textbf{Time, ms}}&{\textbf{Bandwidth, Gb/s }} \\
\hline
1 &	960 & 64 & 64 & 140 & 28.40
\\
\hline
2 &	960 & 64 & 128 & 140 & 28.40
\\
\hline
3 &	960 & 64 & 256 & 140 & 28.40
\\
\hline\hline
4 &	1280 & 64 & 128 & 156 & 25.60
\\
\hline
5 &	1280 & 64 & 256 & 156 & 25.60
\\
\hline\hline
6 &	1600 & 64 & 64 & 156 & 25.61
\\
\hline
7 &	1920 & 128 & 64 & 137 & 29.18
\\
\hline
8 &	2560 & 128 & 64 & 140 & 28.46
\\
\hline
9 &	3200 & 128 & 64 & 124 & 32
\\
\hline
10 &	5120 & 512 & 64 & 124 & 32
\\
\hline
11 & 6400 & 256 & 64 & 109 & 36,57
\\
\hline
12 & 7680 & 512 & 64 & 93 & 42.67
\\
\hline
\end{tabular}
\end{center}
\end{table*}

Table \ref{table2} shows our experimental results.  We observe that, in all data series, an increase in TS by 2 or 4 times does not lead to a change in the array processing speed. At the same time, increasing the size of the WG (which generally increases the overall number of the calculators) increases performance. This can be explained by the fact that, with an increase in the WG size, a larger amount of local instead of global memory is used by Streaming Processors for internal reductions. In addition, since our use case is not computationally complex, some processing elements can be idle at a time, and if there are more tasks, the thread schedulers in the compute units can load them more efficiently.

We also perform experiments with the implementation model described in this section for calculating the minimum on Promela using our counterexample method. We proceed similarly to the approach in 
Section~\ref{limited}, i.e., we find a counterexample of the logical requirement that a successful calculation cannot be reached is a particular program trace. Using this trace, we can find the optimal parameter values for our model. 

\begin{table*}[htbp]
\caption{Experiments with the Minimum Promela model}\label{table3}
\begin{center}
\begin{tabular}{ |p{1cm}|p{2cm}|p{2cm}|p{1cm}|p{1cm}|p{2cm}|p{1cm}|}
\hline
\textbf{N}&\textbf{Processing elements}&{\textbf{Data size}}&{\textbf{WG}}&{\textbf{TS}}&{\textbf{Model time}}&{\textbf{Steps}}\\
\hline\hline
1 & 4 & 16 & 8 & 2& 20 & 1347 
\\
\hline
2 & 4 &	16& 4 & 4 & 24 & 1229 
\\
\hline
3 & 4 &	16& 2 & 4 & 25 & 1365 
\\
\hline
\hline
4 & 64 &	64& 16 & 4 & 36 & 4398  
\\
\hline
5 & 64 &	64& 8 & 8 & 44 & 4101  
\\
\hline
6 & 64 &	64& 4 & 4 & 75 & 4101  
\\
\hline\hline
7 & 64 &	128& 8 & 16 & 76 & 7303  
\\
\hline
8 & 64 &	128& 4 & 16 & 137 & 7493  
\\
\hline
9 & 64 &	128& 4 & 8 & 139 & 7781  
\\
\hline\hline
10 & 64 &	256& 4 & 8 & 271 & 15024  
\\
\hline
11 & 64 &	256& 4 & 4 & 279 & 16151  
\\
\hline
12 & 64 &	256 & 2 & 4  & 295 & 18491
\\
\hline
\end{tabular}
\end{center}
\end{table*}

Table \ref{table3} shows the results with Promela. 
We observe that the WG parameter, as well as in a full-scale experiment, affects the run time more dramatically than TS parameter.  
In the solutions with the best time for a given data size, there were always the largest value of this parameter among other solutions found by the model checker due to more intensive using of local memory as in real experiments. 

Hence, we observe that our Promela-based approach to auto-tuning the Minimum problem implementation in OpenCL demonstrates the same time behaviour of the program performance depending on the tuning parameters WG and TS as the manual approach to finding the optimal parameter configuration.
In our future work, relying on these positive experimental results, we will address scaling up our formal approach to much larger sizes of input data.

\section{Conclusion}\label{conc}

The long-term goal of our research is the development of rigorous, formally-based methods for auto-tuning high-performance parallel programs by relying on the model checking approach, in particular using the technique of counterexample-guided search. The results of auto-tuning are the optimal values of tuning parameters that ensure the best possible performance of the given program on a particular parallel architecture for a particular input data size. As our current target, we address high-performance programs written in OpenCL which is an emerging open standard for programming various kinds of parallel architectures, including multi-core CPU, GPU, and FPGA. Our approach described in this paper is based on developing an abstract model of OpenCL computations and then expressing the model of program execution and optimality properties in the formal language Promela, such this model can be investigated by the popular model checking tool SPIN. 

The main advantage of our logic-based approach to auto-tuning high-performance programs is that is can be applied without involving the particular target execution platform. Other than the existing auto-tuners, our simulation of the execution of OpenCL programs in Promela abstracts from specific computations, keeping the logic of interaction and synchronization of parallel program processes in the target parallel architecture. Due to the formal semantics of the Promela language, the model of a given OpenCL program can be viewed as the formal operational semantics of interaction and synchronization of parallel program processes on a selected platform and device architecture. By varying the parameters and algorithms of interaction between the components of an abstract device, it is possible to define and study specific processor architectures. This makes it possible to search for optimal program settings in the absence of real target processors, while existing auto-tuning frameworks cannot provide such an opportunity.

We plan to further improve our approach by customizing our general auto-tuning method for a specific application program executed on a specific hardware, in addition to computing the minimum on the Nvidia P104-100 processor. In particular, we are going to conduct a case study with matrix multiplication on the most recent Ampere GPU architecture by Nvidia Corp. Knowing the particular specification, for example of the Ampere architecture, we can customize the ratio of the global memory access time to the local memory access time, and the performance time for every function used in the program by considering all primitive operations in the function. We will also take into account specific organisation of communication in the architecture, for example between Ampere units and their processing elements.

In our future work, we will also add the communication time between processes to the model, as well as consider other settings, in particular, the number of work items. In addition, to improve our counterexample method, we are going to try using the never claim pruning from \cite{panizo2017guided}, and finding optimal values from \cite{Ruys03}, which avoids multiple runs of the SPIN verifier in the counterexample method. We also plan to improve the scalability of our counterexample method by modeling a warp-based scheduling \cite{lashgar2013warp}, because due to warping, the number of simultaneously working processing elements becomes smaller, which implies reduction of interleaving non-determinism in the model. Finally, we will work on scaling up our approach to much larger sizes of input data. 

\section*{Acknowledgements}
We are grateful to the anonymous referees of the LOPSTR'22 symposium for their great help in improving the initial, shorter version~\cite{garanina2022model} of this paper. 

\bibliographystyle{fundam}

\bibliography{main.bib}

\end{document}